\documentclass[fleqn,usenatbib]{mnras}

\usepackage{newtxtext,newtxmath}

\usepackage[T1]{fontenc}
\usepackage{ae,aecompl}

\usepackage{graphicx}	
\usepackage{amsmath}	
\usepackage{amssymb}	
\usepackage{multicol}   
\usepackage{bm}         
\usepackage{pdflscape}  
\usepackage{subcaption}
\usepackage{color}
\defcitealias{Park+2018}{Paper~I}
\title[Making top-Heavy IMFs near the GC]{Making top-heavy IMFs from canonical IMFs near the Galactic Centre}

\author[S.-M.~Park, S.~P.~Goodwin and S.~S.~Kim]{
So-Myoung Park,$^{1}$\thanks{E-mail: smp.smpark@gmail.com}
Simon P. Goodwin$^{2,3}$
and Sungsoo S. Kim$^{1,4}\thanks{E-mail: sungsoo.kim@khu.ac.kr}$
\\
$^{1}$School of Space Research, Kyung Hee University, 1732 Deogyeong-daero, Giheung-gu, Yongin-si, Gyeonggi-do 17104, South Korea\\
$^{2}$Department of Physics and Astronomy, University of Sheffield, Sheffield S3 7RH, UK\\
$^{3}$Humanitas College, Kyung Hee University, 1732 Deogyeong-daero, Yongin-si, Giheung-gu, Gyeonggi-do 17104, South Korea\\
$^{4}$Department of Astronomy and Space Science, Kyung Hee University, 1732 Deogyeong-daero, Giheung-gu, Yongin-si, Gyeonggi-do 17104,\\South Korea
}

\date{Accepted XXX. Received YYY; in original form ZZZ}

\pubyear{2020}

\begin{document}
\label{firstpage}
\pagerange{\pageref{firstpage}--\pageref{lastpage}}
\maketitle

\begin{abstract}
We show that dynamical evolution in a strong (Galactic Centre-like) tidal field can create clusters that would appear to have very top-heavy IMFs.
The tidal disruption of single star forming events can leave several bound `clusters' spread along 20~pc of the orbit within 1-2~Myr.
These surviving (sub)clusters tend to contain an over-abundance of massive stars, with low-mass stars tending to be spread along the whole `tidal arm'.
Therefore observing a cluster in a strong tidal field with a top-heavy IMF might well not mean the stars formed with a top-heavy IMF.
\end{abstract}

\begin{keywords}
methods: numerical -- Stars: kinematics and dynamics -- stars: mass function -- Galaxy: centre -- open clusters and associations: general
\end{keywords}



\section{Introduction}

The initial mass function (IMF) of stars is extremely important in many areas of astrophysics as it sets the relative numbers of low, intermediate, and high-mass stars and hence parameters such as chemical yields, mass-to-light ratios, mass-loss rates with time, etc.

The stellar mass function of individual clusters and of the Galactic field has been parameterised in a number of ways, e.g. \citet{Salpeter1955}, \citet{Kroupa2002}, \citet{Chabrier2003}, \citet{Maschberger2013} with an empirical fit to the observed number of stars per unit mass interval to create a `canonical IMF' (against which all other IMFs are compared)\footnote{It should be noted that formally, the IMF is the mass distribution at birth unchanged by stellar or dynamical evolution.
However, it is unclear if a {\em true} IMF can ever be observed.
In this paper we will use the term `IMF' with its generally used meaning of `the mass function of a very young stellar system'.}.

The IMF is often considered to be `universal', i.e. (statistically) the same everywhere \citep[see][for a review]{Bastian+2010}.
But a question of great interest is if the IMF {\em really} is always (statistically) the same, or if it varies with environment, and if it varies how does it vary?
There are various observations that might suggest a statistically significant variation in resolved IMFs (i.e. systems in which we measure the masses of individual stars).
\citet{Dib+2017} finds evidence of significant cluster-to-cluster variations in the IMF among a large sample of young clusters in the Milky Way.
NGC 3603 seems to have a top-heavy IMF \citep{Pang+2013}.

It is worth noting that \citet{Schneider+2018} seem to find that the 30 Dor region in the LMC has an IMF with a flatter slope than the canonical Salpeter slope for stars $>$15~M$_{\odot}$.
It is unclear if this is `top-heavy' or if that is actually the typical slope for high-mass stars as this is the first observation with significant numbers of very high-mass stars to construct a robust IMF at very high-masses.

Of particular interest to us in this paper are observations of IMFs in the Galactic Centre (GC).
\citet{Hosek+2019} apparantly find a top-heavy IMF for the Arches cluster \citep[see also][]{Kim+2006}, and 
the Quintuplet cluster also seems to have a flatter MF than the canonical IMF \citep{Hussmann+2012}.

The question we would like to address is if observing a top-heavy IMF in the GC is evidence of stars having formed with a top-heavy IMF? 
It is quite possible that stars do form with different IMFs in different environments \citep[e.g.][]{Elmegreen+2003,Shadmehri2004,Dib+2007,Hocuk+2010,Dib2014}, but robust observational evidence would be needed to show that they indeed do, and to determine what the IMF actually is.

In this paper we examine the dynamical evolution of star forming regions in a strong (Galactic Centre) tidal field.
We show that it is possible to dynamically form `clusters' from a region with a canonical IMF that would appear to an observer to have a very top-heavy IMFs (as many low-mass stars are `lost' in the general field).  We also show that stars that form within 1 pc of each-other can rapidly (within 1--2 Myr) be separated by $>10$ pc in a strong tidal field.
We caution that determinations of IMFs need to take into account possibly significant dynamical evolution and be sure that all stars that formed together are included in the determination (which may be practically impossible).

\section{Methods}

We run our simulations using Aarseth's {\sc nbody6} code \citep{Aarseth1999} with full (non-truncated) tidal forces \citep{Kim+2000}.
We simulate clumpy and substructured (fractal) star clusters at distances of 30 and 100~pc, respectively, from the Galactic Centre (GC) following \citet[hereafter \citetalias{Park+2018}]{Park+2018}.

Based on the results of \citet{Park+2018} we have chosen initial conditions that we know do not survive as a single cluster, but are shredded by the tidal field but retain significant substructures that an observer at a later time might think are individual clusters.

We evolve the star clusters for $\sim$2~Myr, the (minimum) age of star clusters that we can observe near the GC, e.g. Arches (2-4~Myr; \citealt{Martins+2008,Clark+2018a}) and Quintuplet (3-4~Myr; \citealt{Liermann+2012,Clark+2018b}) clusters.

\subsection{Tidal field}

We model the Galactic potential using Oort's $A$ and $B$ constants \citep{Oort1927}.
We assume that (a) the Galactic enclosed mass profile at 30 and 100~pc from the GC follows a power law \citep{Kim+1999}, and (b) star clusters are moving in a circular orbit so the Galactic potential is constant with time at a particular Galactocentric distance.
We use the Galactic enclosed mass from \citet[see their Fig.14]{Launhardt+2002}.
For a more realistic Galactic potential, we also consider the effective potential.
The differential gravitational potential and the centrifugal potential are included.
Full details can be found in \citetalias{Park+2018}.

\subsection{Cluster mass and IMF}

We pick a mass to match the total photometric mass of the Arches cluster of $\sim$2.0~$\times~ 10^{4}~{\rm M}_{\odot}$ from the best fit model for the Arches cluster from \citet{Kim+2000}.

We randomly select stellar masses from the \citet{Maschberger2013} IMF from 0.01 to 100~M$_{\odot}$.
The probability density function (PDF) of \citet{Maschberger2013} is
\begin{equation}
	\label{eq2.1}
	\begin{aligned}
		p(m) \propto \bigg(\frac{m}{\mu}\bigg)^{-\alpha} \Bigg(1 + \bigg(\frac{m}{\mu} \bigg)^{1-\alpha} \Bigg)^{-\beta},
	\end{aligned}
\end{equation}
where $\mu$ is the average stellar mass, $\alpha$ is the Salpeter power-law index for high mass stars, and $\beta$ is the power-law index for low mass stars.
This PDF combines the log-normal approximation for the IMF derived by \citet{Chabrier2003} with \citet{Salpeter1955} power-law slope for stars with mass > 1~M$_{\odot}$.
Here, we use $\mu = 0.2~{\rm M}_{\odot}$, $\alpha = 2.3$, and $\beta = 1.4$ within the mass range of stars $m_{\rm low}=0.01$ to $m_{\rm up}=100~{\rm M}_{\odot}$ and set the total number of stars to be $N=31000$.
This gives a total cluster mass $\sim$2.0~$\times~10^{4}~{\rm M}_{\odot}$.

Masses are initially distributed at random in the clusters so there is no primordial mass segregation.
To measure the initial mass segregation of star-forming regions quantitatively, we use the  $\Lambda_{\rm MSR}$ method \citep{Allison+2009a}
The $\Lambda_{\rm MSR}$ measures the degree of mass segregation.
When the $\Lambda_{\rm MSR}$ is significantly $>1$ then the star clusters are mass segregated.
If it is $\sim$1, it means there is no mass segregation.
In Table~\ref{table2}, we calculate the $\Lambda_{\rm MSR}$ for the 10 most massive stars at 0~Myr of each model.
The $\Lambda_{\rm MSR}$ of each model is $\sim$1 so initial star-forming regions are not mass segregated.

\subsection{Initial position and velocity structures}

\begin{table}
	\centering
	\caption{A summary of initial conditions. $R_{\rm GC}$ is the Galactocentric radius, $R_{\rm t}$ is the nominal tidal radius, $R_{\rm c}$ is the total (outer) cluster radius, and $V$ is the virial ratio.}
	\label{table1}
	\begin{tabular}{ccccc}
		\hline 
		Model & $R_{\rm GC}$ & $R_{\rm t}$ & $R_{\rm c}$ & $V$ \\ 
		\hline 
		Fractal & 30~pc & $\sim$2.0~pc & $\sim$2.0~pc & 0.3\\
                & 100~pc & $\sim$3.0~pc & $\sim$3.0~pc &   \\ 
		\hline
	\end{tabular}
\end{table}

\begin{table}
	\centering
	\caption{The initial $Q$-parameters and $\Lambda_{\rm MSR} \pm 1 \sigma$ of each model.}
	\label{table2}
	\begin{tabular}{ccc}
		\hline 
		Model & $\Lambda_{\rm MSR}$ & $Q$-parameter \\ 
		\hline 
		{\rm A} & $1.28\pm0.16$ & 0.47 \\
        {\rm B} & $1.47\pm0.22$ & 0.56 \\
        {\rm C} & $0.96\pm0.15$ & 0.51 \\
        {\rm D} & $0.92\pm0.13$ & 0.50 \\
        {\rm E} & $0.91\pm0.12$ & 0.47 \\
		\hline
	\end{tabular}
\end{table}

\begin{figure*}
	\centering
	\includegraphics[width=0.9\columnwidth,angle=90]{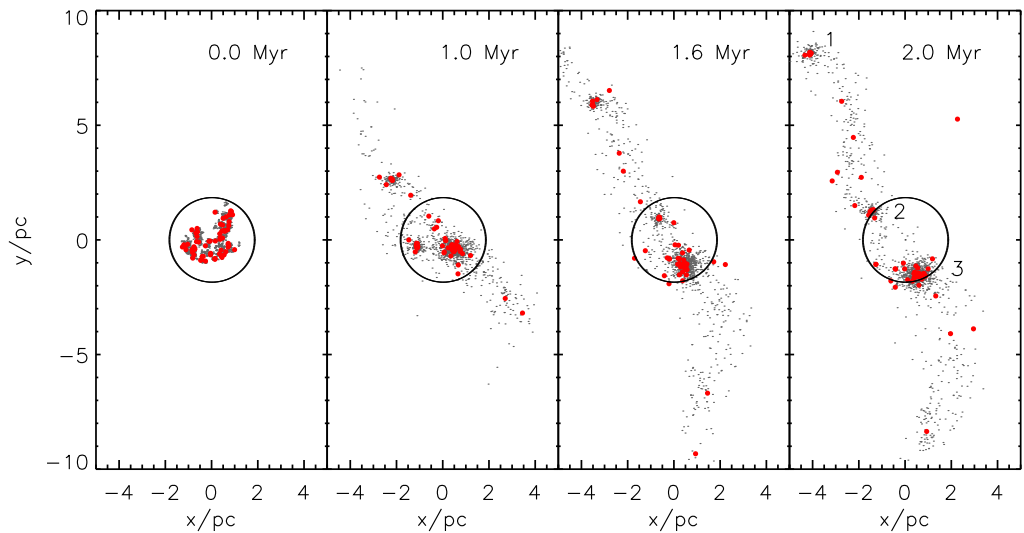}
    \caption{The evolution of a cool ($V=0.3$) fractal star cluster at 30~pc from the GC.
    Grey dots are stars massive than 2~M$_{\odot}$ and big red dots are those more massive than 20~M$_{\odot}$.
    The black circle in the middle of each panel is the initial nominal tidal radius.
    In the final panel are the `cluster IDs' (see text).} 
    \label{fig01}
\end{figure*}

We choose `fractal' initial conditions (ICs) for an initial distribution of stars in a cluster to model a complex initial distribution.
Stars seem often to form in clumpy and filamentary substructures \citep{Konyves+2015,Lu+2018,Parker2018,Dib+2019}.
Note that we are not claiming that this is a match to how GMCs actually produce stars, rather it is a simple method for making (sub)structured initial conditions that are closer to reality that the classic Plummer sphere.

Fractals are constructed following \citet{Goodwin+2004} with a fractal dimension $D = 2.0$ (moderately substructured fractals).
To make a fractal initial distribution, a box is defined with a `parent' star at the centre of the box.
The box is then divided into sub-boxes, and each sub-box contains a `child' star at its centre.
The fractal dimension $D$ determines which child stars become a parent star of the next generation.
Velocities are inherited from parents with an extra random component that reduces in magnitude with depth in the fractal.
This means that near neighbours tend to have similar velocities, while distant neighbours can have very different velocities.
See \citet{Goodwin+2004} for full details.
Note that masses are assigned randomly, so there is no correlation between mass and velocity initially.

To measure the amount of substructure to compare with real star forming regions we use the $Q$-parameter \citep{Cartwright+2004}.
In Table~\ref{table2}, we measure the initial $Q$-parameters of each model and the values are $\sim$0.55, which are similar to many low/intermediate mass clusters in the Solar neighbourhood \citep{Cartwright+2004}, and to the distributions of cores in e.g. Orion B \citep{Konyves+2020}.
However, the $Q$-parameter of young regions can change rapidly due to dynamics \citep{Allison+2009b,Parker+2014}, or the massive star-forming region ``may" start already with higher $Q \sim 0.7-0.75$ value than those found in low/intermediate mass regions \citep{Dib+2019}.
So it is unclear if local regions are a good analogue to star formation in the GC \citep[e.g.][]{Barnes+2019}.
But we would argue that it is as reasonable a starting point as any.

Because the dynamical evolution of fractals is highly stochastic, statistically identical initial conditions can evolve very differently \citep{Allison+2010}.
Therefore we pick a number of different realisations from \citetalias{Park+2018} at 30 and 100~pc from the GC.

\subsection{Initial cluster size and internal energy}

In \citetalias{Park+2018} we were interested in what initial conditions could survive as a single massive cluster.
In that paper we found that if the initial size of a fractal region is similar or larger than the `nominal tidal radius' (the tidal radius of the region if it were a point mass), then regions could be shredded but contain significant surviving substructure.

Based on \citetalias{Park+2018} we pick initially cool fractal regions with a isolated virial ratio ($V$) of 0.3, and a radius of 2~pc at 30~pc from the GC, and 3~pc at 100~pc from the GC (see table \ref{table1}).

\subsection{Stellar evolution and binarity}

We do not include stellar evolution in our simulations as it is not important for any but the most extremely massive ($>$100~M$_\odot$) stars in the $\sim$2~Myrs of our simulations.

We do not include binaries mainly for practical reasons.
Including binaries adds significant computational expense, and opens-up a huge area of parameter space (binary mass ratios and separation distributions).
Binaries can alter dynamics significantly acting as a heat sink or source \citep[e.g.][]{Heggie1974,Hills+1974,Hills1990}, and the formation and destruction of massive multiple systems can alter the dynamics of whole systems \citep[e.g.][]{Allison+2011}.
However, we argue that our simulations show a plausible mechanism for producing apparently top-heavy IMFs via dynamics which the addition of binaries would probably not change significantly.

\bigskip

We simulate $\sim$2.0~$\times~10^{4}$~M$_{\odot}$ ($N=31000$) clusters for 2~Myr in a realistic strong tidal field 30 and 100~pc from the GC and the tidal radii at those positions are $\sim$2 and 3~pc, respectively.
Our initial distributions are cool ($V=0.3$) and $D=2.0$ fractal distributions.
The initial size of fractal star clusters is similar to their tidal radii.
A detailed summary of the initial conditions that we use is given in tables~\ref{table1} and \ref{table2}.

\section{Results}

\begin{figure}
	\centering
	\includegraphics[width=0.9\columnwidth]{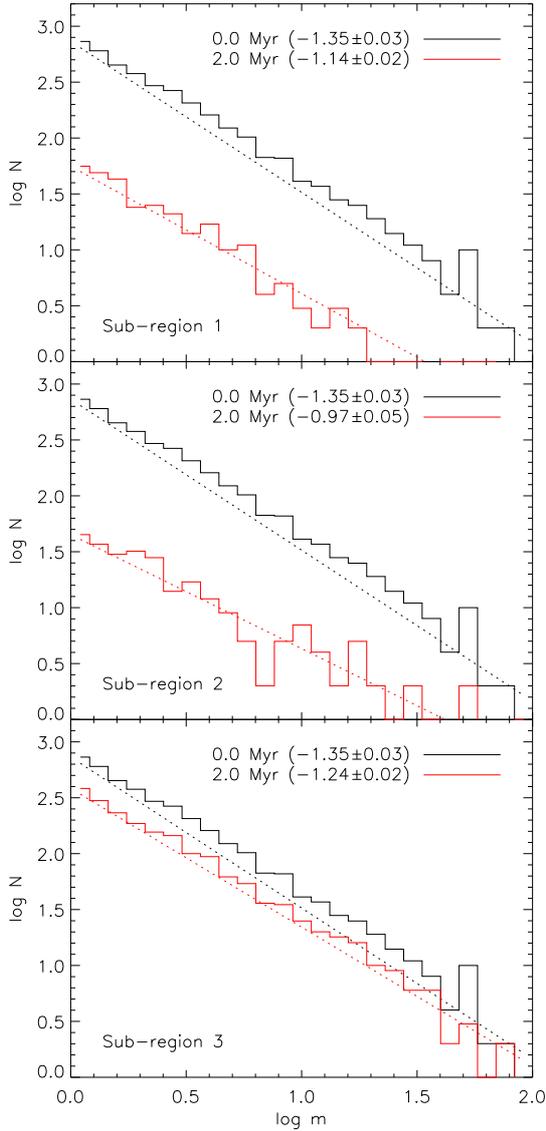}
    \caption{The evolution of the MF slope over 2~Myr for each sub-region in Fig.~\ref{fig01}.
    The black and red histograms are the MFs at 0 and 2~Myr, respectively.    
    The dotted lines are the best-fit power-law slopes, with the 
    numbers in a parenthesis the best-fit power-law index.}
    \label{fig02}
\end{figure}

In \citetalias{Park+2018} we showed that regions that fill, or over-flow their nominal tidal radii will rapidly be `shredded' by a strong tidal field.
In that paper we were interested in those regions that survived as a single cluster, but in this paper we will look in more detail at regions that are shredded by the tidal field.

\subsection{Regions at 30~pc from the GC}

\begin{figure*}
	\center
	\includegraphics[width=0.9\columnwidth,angle=90]{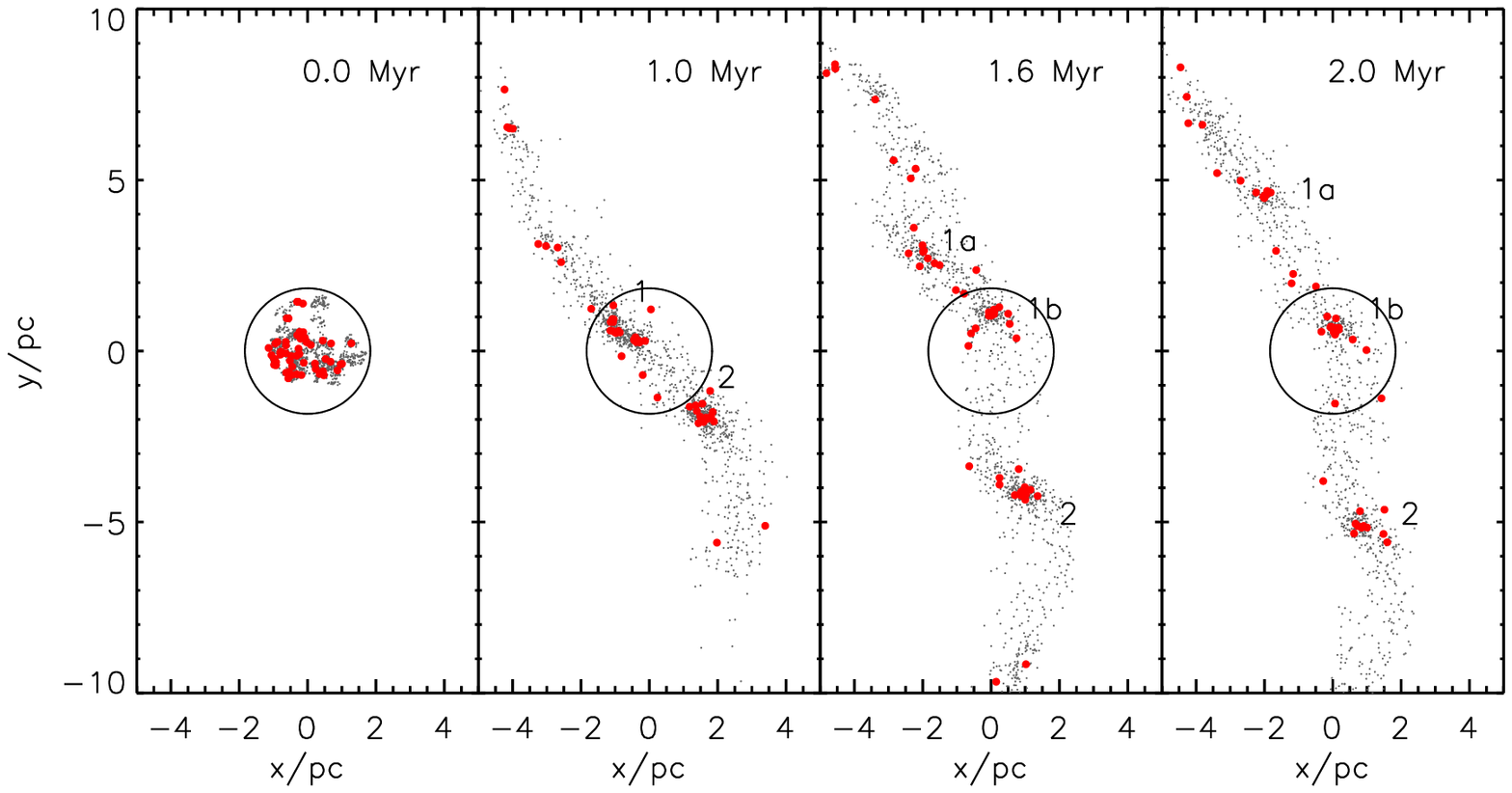}
	\includegraphics[width=0.9\columnwidth,angle=90]{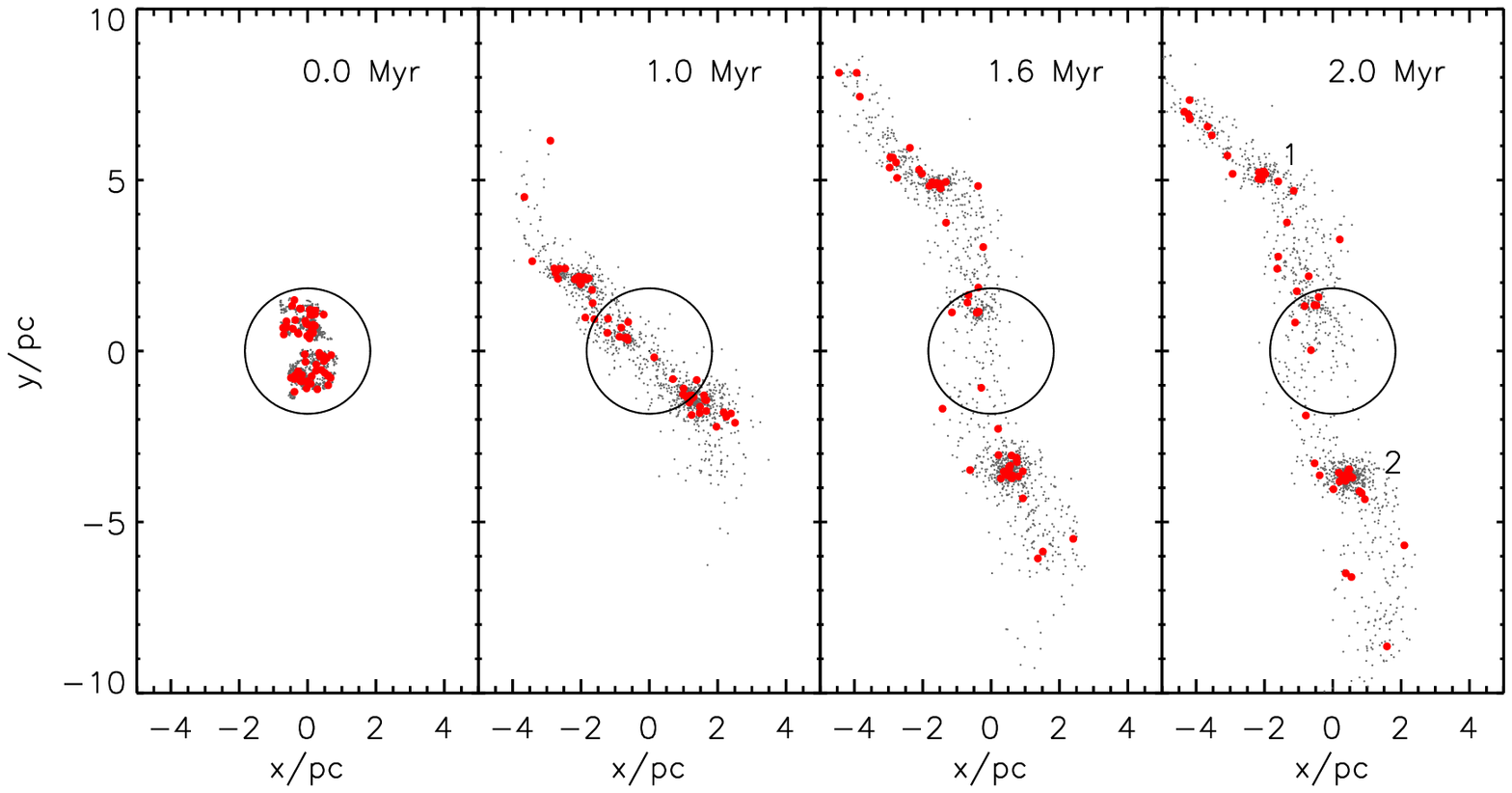}
    \caption{The evolution of two other realisations of a cool ($V=0.3$) fractal star cluster at 30~pc from the GC at 0, 1, 1.6 and 2~Myr.
    Grey dots are stars massive than 2~M$_{\odot}$ and red big dots are those massive than 20~M$_{\odot}$.
    The black circle in the middle of each panel is the nominal tidal radius.  In the final panels are the `cluster IDs' (see text).}
    \label{fig03}
\end{figure*}

\begin{figure}
\centering
    \includegraphics[width=0.9\columnwidth]{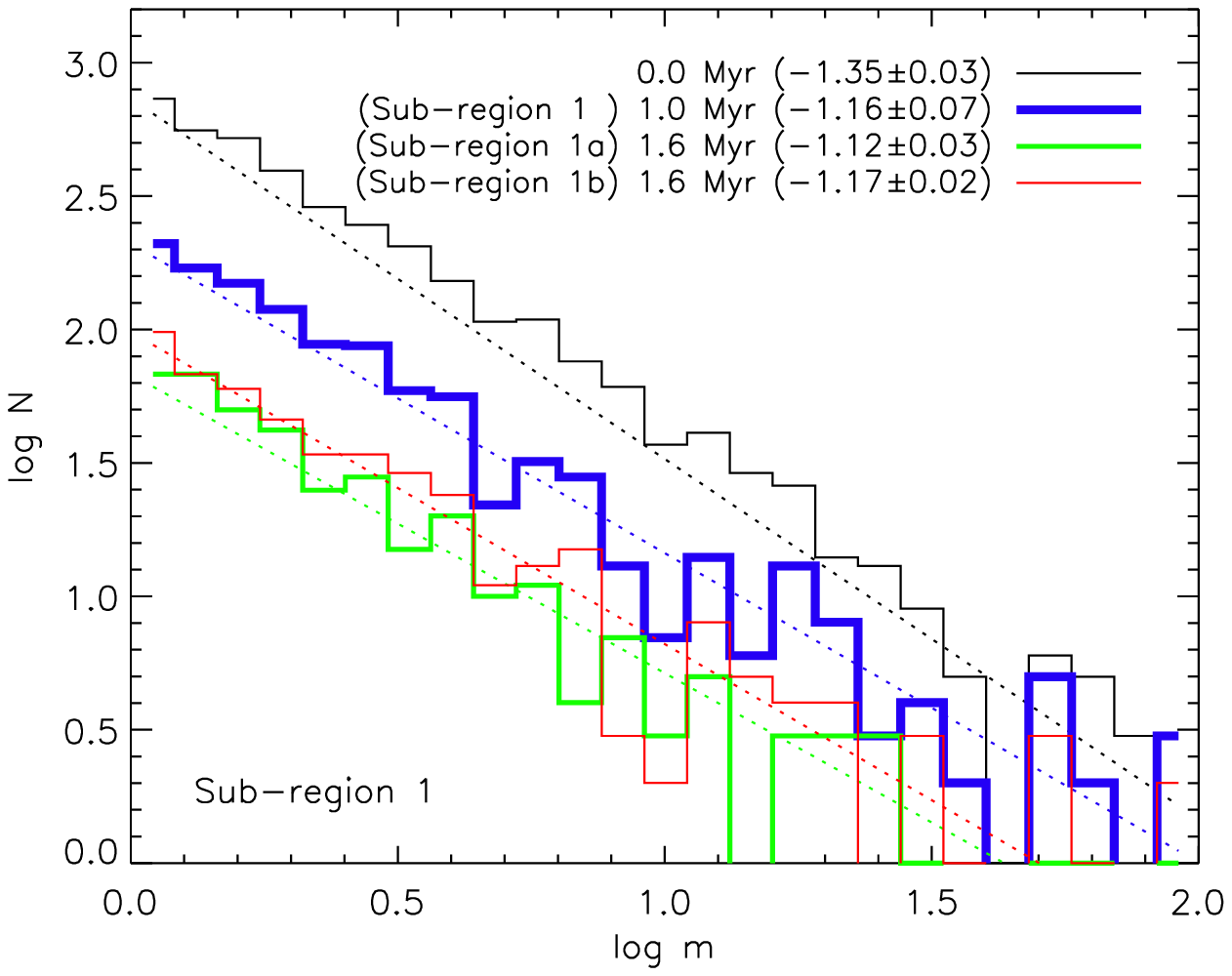}
    \caption{The variation of slope of MF of sub-region 1 between 1 and 1.6~Myr.
    Black and blue histograms are MF at 0 and 1~Myr.
    Green and red histograms are MF of sub-region 1a and 1b at 1.6~Myr.
    }
    \label{fig04}
\end{figure}

In Fig.~\ref{fig01} we show the evolution of a particular cool ($V=0.3$), fractal ($D=2$) region at 30~pc from the GC.
The red dots are the  massive stars $> 20$~M$_\odot$ that would be relatively easy to observe, and the grey dots are all other stars (mostly $<$0.5~M$_\odot$ and so essentially unobservable in the GC).
The black circle shows the nominal tidal radius of the region (the tidal radius if the region was a point mass).
Note that this region has a completely canonical global IMF.

The region evolves in time from left to right in Fig.~\ref{fig01}.
By 1~Myr (second panel) it has significantly elongated due to the strong tidal field, and by 2~Myr (right most panel) the stars are spread over more than 20~pc around the orbit.

Let us consider what an observer would see if observing this region after 2~Myr of dynamical evolution near the GC.
Note that our view is down onto the plane of the orbit, so from above (and outside) the plane of the Galaxy, so this view does not match what an observer on the Earth would see (we will return to reconstructing what an observer on the Earth would see in a later paper).

Firstly, much of the low-mass population (grey dots) has been spread through the tidal arms.
If we consider the total stellar population (red and grey points) the tidal structure is very obvious.
However, in the GC these (usually very faint) low-mass stars would be `lost' in a confusion of other stars in the GC as well as significant foreground and background contamination.
Therefore, an observer would first only easily identify the high-mass stars (the red dots).

One thing that would stand-out to an observer would be a significant `cluster' (at around (0,-1)~pc) which we have labelled `3'.
There are also two other, much less massive, clusters visible at (-4,8)~pc (labelled `1'), and at (-2,1)~pc (labelled `2').
There are also a few `isolated' massive stars which would be obviously young and massive, but it is unclear if they would be associated with the three `clusters'.

Because these `clusters' are spatially distinct it would be easy to think that they formed as different `units' of star formation.
As they are the same age they would seem related in some way, but it would not be obvious that the stars currently in cluster `1' all formed with roughly 1~pc of all the stars now in cluster `3'.  

Any attempt to measure an IMF will concentrate on the regions identified as `clusters' which will (a) not know that the three `clusters' were all part of the same star formation event, and (b) ignore significant numbers of low-mass stars which are either unobservable or no longer obviously associated with the `clusters'.

The cause of the evolution we see in Fig.~\ref{fig01} is very easy to understand.
The strong tidal field is essentially a shear force on the stars in the region.
This force causes the global `elongation' of the region along its orbit giving the large-scale classic tidal structure which appears.  

Because the initial region is clumpy (fractal) it has regions of high-density that are able to survive as bound sub-units.
An initial phase of violent relaxation erases small-scale structure and causes sub-regions that survive to become roughly spherical.
Generally, sub-units will be much more likely to be bound if they contain massive stars as this gives them more mass and makes them more bound.
As the sub-units are bound they will undergo internal dynamical evolution which will cause them to both mass-segregate and lose some of their lower-mass stars.  

Therefore we have two effects that act to mean that bound sub-units will have `too many' massive stars: a bias that regions with  massive stars are more likely to be bound, and then that these regions will tend to lose low-mass stars preferentially.

Hereafter, we use the term `sub-region' instead of `clusters' to distinguish the apparent `clusters' in our simulations from `genuine' clusters which still contain (almost) all the stars that formed together at the same time.

In Fig.~\ref{fig02}, we measure the slope of the mass function (MF) above $1$~M$_\odot$ at 2~Myr of each sub-region in Fig.~\ref{fig01}.
This MF would almost certainly be called `the IMF of the cluster' (it is the MF of young stars in a locally bound region).
In what follows we will talk about the `apparent IMF', or aIMF as we are observing the current MF in a particular sub-region which {\em we} know may not be representative of the actual IMF.

The black histogram is the (true) IMF of the whole region, and is (by design) a canonical IMF.
The red histograms are the aIMFs that would be measured if just observing sub-regions 1 (top panel), 2 (middle panel), and 3 (bottom panel).
In each case we find the best-fit slope to the observed aIMF.
In the most massive sub-region 3 (bottom panel) the aIMF slope is $-1.24$ which is somewhat flatter than the canonical $-1.35$ of our true IMF, and sub-regions 1 and 2 have aIMF slopes of $-1.14$ and $-0.97$ respectively, which are very significantly top-heavy.
We would argue that an observer apparently finding a 2~Myr old cluster 10~pc away from any other recent star formation with an aIMF slope of -1.1 or -1 would consider this to be a very top-heavy star formation event.

Fig.~\ref{fig03} shows the evolution of two other realisations of a clumpy region at 30~pc from the GC.
The details of the evolution of each of these regions are different to that shown in Fig.~\ref{fig01}, but the overall picture is the same.
In both cases the regions are spread over 20~pc in extent by the tidal field with sub-regions containing `too many' massive stars remaining.
In the top panel of Fig.~\ref{fig03} the sub-regions labelled `1a', `1b', and `2' would be measure to have aIMF slopes of -1.10$\pm$0.02, -1.09$\pm$0.03, and -1.05$\pm$0.3 respectively at 2~Myr.
And in the lower panel of Fig.~\ref{fig03} the sub-regions labelled `1' and `2' would have aIMF slopes of -1.03$\pm$0.05 and -1.23$\pm$0.02, respectively at 2~Myr.

\subsection{The evolution of sub-regions}

Sub-regions can evolve in quite complex ways.
We can see the significant evolution of a sub-region between 1 and 1.6~Myr, in the second and third panel of the top panel of Fig.~\ref{fig03}.

At 1~Myr (the second panel) there are two sub-regions an observer might distinguish, one at (-1,1)~pc (labelled `1'), and another at (2,-2)~pc which are roughly 5~pc apart.
But by 1.6~Myr (the third panel) there are 3 sub-regions.
The lower sub-region at 1~Myr has moved from (2,-2)~pc to (1,-5)~pc, but the top sub-region has split into two.
Now labelled `1a' and `1b' we have a sub-region at (0,1)~pc, and another at (-2,5)~pc.

Fig.~\ref{fig04} shows the aIMFs that we would measure for sub-region at different times.
The blue line shows the aIMF of sub-region 1 at 1~Myr which has an apparently top-heavy slope of -1.16.
At 1~Myr the sub-region is clearly elongated, and knowing that it splits it is possible to see two distinct `halves' to the sub-region.
By 1.6~Myr there are two separate sub-regions, both with top-heavy aIMF slopes of a little less than -1.35, but now separated by approximately 2~pc.
Again, it is unclear if an observer would consider if two `clusters' separated by 2~pc were actually a single `cluster' just 0.6~Myr previously.

\begin{figure*}
	\center
	\includegraphics[width=0.9\columnwidth,angle=90]{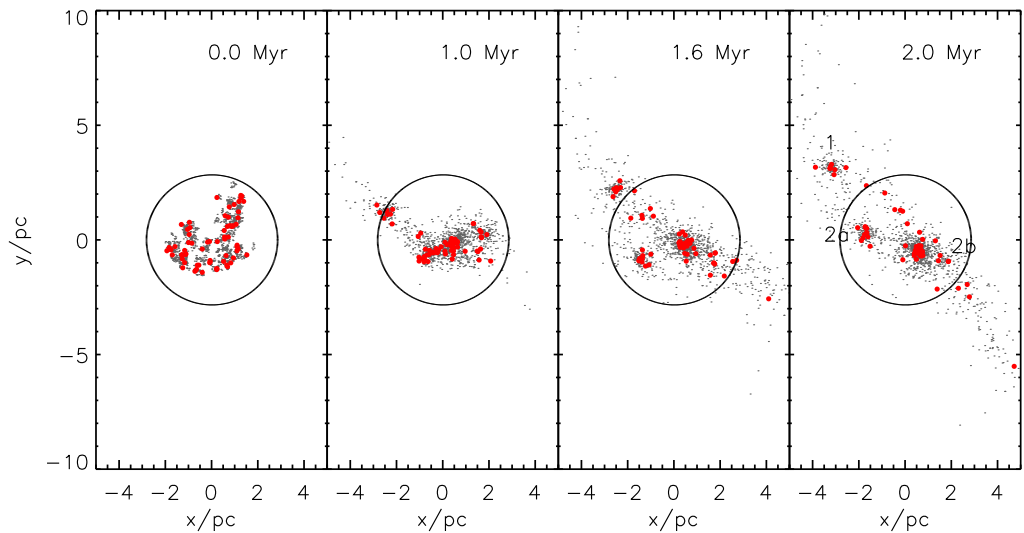}
    \caption{The evolution of a cool ($V=0.3$) fractal star cluster at 100~pc from the GC.
    Grey dots are stars massive than 2~M$_{\odot}$ and red big dots are those massive than 20~M$_{\odot}$.
    The black circle in the middle of each panel is a nominal tidal radius.
    In the final panel are the `cluster IDs' (see text).}
    \label{fig05}
\end{figure*}

\subsection{A typical region 100~pc from the GC}

The strength of the tidal field at 100~pc from the GC is quite significantly less than that at 30~pc (although still strong by any usual definition of tidal field strength), and so the shredding of initial regions is somewhat less extreme.

In Fig.~\ref{fig05} we show the evolution of a cool ($V=0.3$) fractal star cluster at 100~pc from the GC.
We use the same fractal distribution as that illustrated in Fig.~\ref{fig01}, but scale it to be slightly bigger so that it just fills its nominal tidal radius at 100~pc from the GC (previously it just filled its nominal tidal radius at 30~pc from the GC).

At 0~Myr (the leftmost panel of Fig.~\ref{fig05}) there is one clumpy and substructured region (a slightly larger version of the first panel of Fig.~\ref{fig01}) just filling the nominal tidal radius of $\sim 3$~pc.

Over the first 1~Myr the region undergoes violent relaxation whilst also under the influence of a strong tidal field.
As the tidal field is somewhat weaker its effect is somewhat less than at 30~pc from the GC and most stars are still within the nominal tidal radius (cf. the second panel of Fig.~\ref{fig01}).
The region appears to have two main clumps by 1~Myr: a smaller sub-region at (0,0)~pc, and a large, elongated region at the centre of the frame.
The slightly weaker tidal field has had less effect in 1~Myr than at 30~pc from the GC so there are no obvious tidal features yet.

By 1.6~Myr (third panel) tidal arms have become apparent in the low-mass stars, and the large elongated sub-cluster present at 1~Myr has divided into two distinct regions a little over 1~pc apart.
And by 2~Myr (forth panel) there are three distinct sub-regions which are labelled `1', `2a', and `2b'.

Because of the weaker tidal field at 100~pc from the GC all three sub-regions are still relatively close together (`2a' and `2b' separated by $\sim 2$~pc, with `1' around 2-3~pc away).
It is probable that these would be considered as related star formation events (unlike in Fig.~\ref{fig01}).

\begin{figure}
	\center
	\includegraphics[width=0.9\columnwidth]{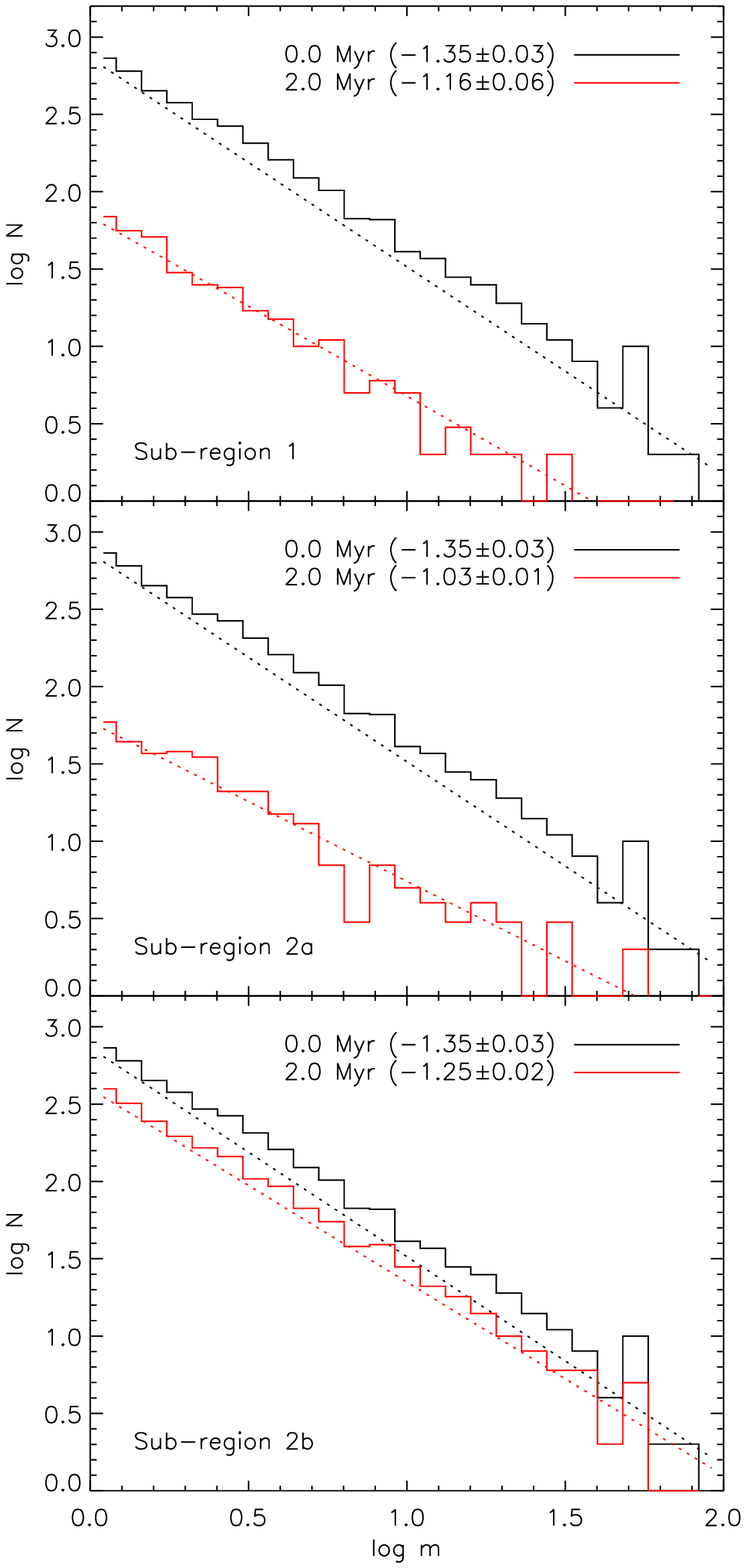}
    \caption{The evolution of the MF slope over 2~Myr of each sub-region in Fig.~\ref{fig05}.
    Black and red histograms are the MF at 0 and 2~Myr, respectively.
    The dotted lines are the best-fit power-law slopes, with the 
    numbers in a parenthesis the best-fit power-law index.}
    \label{fig06}
\end{figure}

Fig.~\ref{fig06} shows the aIMFs that would be measured for the three sub-regions at 2~Myr in Fig.~\ref{fig05}, and shows that all three sub-regions would appear top-heavy with slopes of -1.16, -1.03, and -1.25 respectively. 
Adding these regions together as part of a single star formation event would still result in a seemingly top-heavy aIMF.
From the last panel of Fig.~\ref{fig05} the reason for this is clear as a significant number of low-mass stars have been spread along the tidal arms (visible to us as the grey points, but lost in confusion to an observer).

\section{Conclusions}

Using the {\sc NBODY6} \citep{Aarseth1999} code, we evolve cool ($V=0.3$) fractal star clusters for 2~Myr in a strong tidal field at 30 and 100~pc from the Galactic Centre (GC).
Their initial mass is $ \sim 2 \times 10^{4}$~M$_{\odot}$ and their size is similar to their tidal radii $\sim 2$~pc at 30~pc from the GC and $\sim 3$~pc at 100~pc from the GC (i.e. these are all stars born in the same star-forming region at the same time).
Masses are drawn from a \citet{Maschberger2013} canonical IMF and distributed randomly in the initial region (i.e. no initial mass segregation).

We show that these clumpy star-forming regions can be rapidly `shredded' by the strong tidal field near the GC, but that initially dense and massive sub-regions can survive.
The regions that survive are likely to have an relative over-abundance of massive stars simply because more massive stars have a deeper potential well, and sub-regions containing several high-mass stars are more massive.

Whilst one might argue with the details of the fractal initial conditions, the underlying physical mechanisms behind the process do not depend on the details of the initial conditions.  It simply requires that clumpy/structured initial conditions are not sufficiently bound to survive in the strong tidal field.

In the very strong tidal field at 30~pc from the GC, the initially compact ($\sim 2$~pc) star forming region is extended along tidal arms stretching 20~pc within 2~Myr.  At 100~pc from the GC, the evolution is less extreme due to the weaker tidal field, so surviving sub-regions are closer together after 2~Myr than for 30~pc from the GC.

The surviving sub-regions can relax and look like classic spherical `clusters' several pc apart that might not be thought to have the same origin.
A hypothetical observer would probably consider there to be two or three clusters and would be unable to see the dispersed low-mass component due to confusion and its inherent faintness.

If the mass functions (MFs) of these `clusters' are measured they are often found to be very top-heavy with slopes of -1.2 to -1.0 (cf. the canonical slope of -1.35).
It is quite possible that a hypothetical observer would then report an observation of a `cluster with a top-heavy IMF'.

Therefore {\em measuring a top-heavy IMF in a `cluster' in a strong tidal field does not mean that stars formed with a top-heavy IMF}.
If one were to observe a cluster in the GC with a canonical IMF that might give confidence that one is seeing the majority of stars that formed together, however observing a top-heavy IMF might just be an indication that the `cluster' has undergone significant dynamical evolution in a strong tidal field (the only way to be sure the IMF was truly top-heavy would be to be sure that the surrounding few pc contains very few low-mass stars of the same age).

Therefore the observation by \citet{Hosek+2019} that the Arches cluster close to the GC has a top-heavy IMF might be telling us that those stars formed with a top-heavy IMF, or that they formed with a canonical IMF and dynamical evolution has stripped significant numbers of lower-mass stars.
A search within 20~pc of the Arches for low-mass stars of the same age as the Arches would be required to say for sure.

\section*{Acknowledgements}

Thanks to the referee for their comments which helped us clarify the explanations.
This work was supported by the National Research Foundation grant funded by the Ministry of Science and ICT of Korea (NRF-2014R1A2A1A11052367).
This work was also supported by the BK21 plus program through the National Research Foundation (NRF) funded by the Ministry of Education of Korea.
SPG thanks Kyung Hee University for their financial support and hospitality.









\bsp	
\label{lastpage}
\end{document}